\documentclass[12pt]{article}
\usepackage{amsfonts}

\def\be{\begin{equation}}
\def\ee{\end{equation}}
\def\bea{\begin{eqnarray}}
\def\eea{\end{eqnarray}}

\newcommand{\sect}[1]{\setcounter{equation}{0}\section{#1}}
\newcommand{\subsect}[1]{\subsection{#1}}

\newcommand{\bq}{\mathbf{q}}
\newcommand{\bp}{\mathbf{p}}
\newcommand{\rsl}{\mathfrak{sl}}

\def\kk{K}
 
\def\1{\'{\i}}

\def\k{{\kappa}}
\def\y{{q}}
\def\ji{{\tilde q}}

\def\bjj{{\bf J}}
\def \lkc{{\cal L}^{\rm qg} }
\def \llkc{{\cal L}^{\rm g}}

 \def \g{{\rm g}}

\newcommand{\pji}{{\tilde p}}

\def\dd{{\rm d}}
\def\>#1{{\mathbf#1}}                 
\def\RR{\mathbb{R}}

\def\m{\mu}

\def\la{\lambda}

\begin{document}

\thispagestyle{empty}

\hfill \

\ 
\vspace{0.5cm}

\noindent
 {\large{\bf {Maximal superintegrability of  the generalized  
\vspace{0.2cm}
 \\
  Kepler--Coulomb system on $N$-dimensional curved spaces
 \vspace{0.2cm}
\\
  }} }

\bigskip

\medskip

\begin{center}
\'Angel Ballesteros$^a$ and Francisco J. Herranz$^b$
\end{center}

\noindent
{$^a$ Departamento de F\'\i sica, Facultad de Ciencias, Universidad de Burgos,
09001 Burgos, Spain\\ ~~E-mail: angelb@ubu.es\\[10pt]
$^b$ Departamento  de F\'\i sica,  Escuela Polit\'ecnica Superior, Universidad de Burgos,
09001 Burgos, Spain \\ ~~E-mail: fjherranz@ubu.es 
}

  \medskip

\begin{abstract} 
\noindent
The superposition of the Kepler--Coulomb potential on the 3D Euclidean space with {\em three} centrifugal  terms has recently been shown to be maximally superintegrable  [Verrier P E and Evans N W 2008 J. Math. Phys. {\bf 49} 022902] by finding an additional (hidden) integral of motion which is {\em quartic} in the momenta. In this paper we present the generalization of this result to the $N$D spherical, hyperbolic and Euclidean spaces by making use of a unified symmetry approach that makes use of the curvature parameter.  The   resulting Hamiltonian,  formed by the  (curved) Kepler--Coulomb potential together with  $N$ centrifugal   terms, is shown to be endowed with $2N-1$  functionally independent  integrals of the motion: one of them is quartic and the remaining ones are quadratic. The transition from the proper Kepler--Coulomb potential, with its associated   {\em quadratic} Laplace--Runge--Lenz $N$-vector, to the generalized system is fully described. The role of spherical, nonlinear (cubic), and coalgebra symmetries in all these systems is highlighted.

\end{abstract}

\bigskip\bigskip\bigskip\bigskip



\newpage


\sect{Introduction}

The Kepler--Coulomb (KC) potential on Riemannian spaces of constant curvature 
   was already studied by Lipschitz and Killing   in the 19th century, and later rediscovered by Schr\"odinger~\cite{Sc40} (see~\cite{Sh05} for a detailed discussion). In terms of a geodesic radial distance $r$  between the particle and the origin of the space, the KC potential  on the $N$-dimensional ($N$D) spherical ${\mathbb S}^N$, Euclidean ${\mathbb E}^N$ and hyperbolic ${\mathbb H}^N$ spaces reads (see, e.g., \cite{BH07,kiev,CRS} and references therein)
\be
-\frac{\kk}{\tan r}   \quad   {\rm on}\quad  {\mathbb S}^N;\qquad 
-\frac{\kk}{ r }\quad  
  {\rm on}\quad {\mathbb E}^N; \qquad 
- \frac{\kk}{\tanh r }\quad  {\rm on}\quad  {\mathbb H}^N.
\label{ab}
\ee

In this paper we shall deal with the integrability properties of the so-called  {\em ND generalized KC system}, 
which is defined as the superposition of the (curved) KC potential with  $N$ `centrifugal' terms. In the $N$D Euclidean space $\mathbb  E^N$ such  a system reads
\be
{\cal H}=\frac 12 \bp^2-\frac{\kk}{\sqrt{\bq^2}}+\sum_{i=1}^N \frac {b_i}{q_i^2}
\label{aa}
\ee
where $\kk $ and $b_i\ (i=1,\dots,N)$ are real constants.
 This system was known to be {\em quasi-maximally superintegrable}~\cite{BH07}   in the Liouville sense~\cite{AKN97}, since a set of $2N-2$ functionally independent {\em quadratic} integrals of motion (including the Hamiltonian) were explicitly known.   In fact, in the remarkable classification on superintegrable systems on $ \mathbb E^3$ by Evans~\cite{Ev90},   this Hamiltonian was called `weakly' or {\em minimally superintegrable} since it had one integral of motion more (four)  than the necessary number to be  completely integrable   (three), but one less than the maximum possible number of independent integrals for a 3D system (five). 
 
In contrast, it was also well known that when at least {\em one}  of the centrifugal terms vanishes (we shall call this case the {\em quasi-generalized KC system}), the resulting Hamiltonian turns out to be {\em maximally superintegrable} in arbitrary dimension  since a maximal set of $2N-1$ functionally  independent and {\em quadratic} integrals of the motion is explicitly known (see~\cite{BH07,Ev90,Schrodingerdual,Miguel,Williamsx} and references therein). Moreover, such maximal superintegrability of the quasi-generalized KC system has also  been proven for  the spherical and hyperbolic spaces~\cite{RS, PogosClass1, PogosClass2}  as well as  for  the
Minkowskian and (anti-)de Sitter spacetimes~\cite{kiev, vulpiPAN}.

Nevertheless, in a recent work Verrier and Evans~\cite{Verrier} have shown that the   generalized KC system on ${\mathbb E}^3$ ({\em i.e.}, the superposition of the 3D KC potential with three centrifugal  terms) is {\em maximally suprintegrable}, but the additional  integral of motion is {\em quartic} in the momenta. The aim of this paper is  to  show that this result holds for  an arbitrary dimension $N$ and, moreover, that
the generalized KC system is also maximally superintegrable on the $N$D curved  Riemannian spaces of constant curvature: the spherical ${\mathbb S}^N$ and hyperbolic ${\mathbb H}^N$ spaces. In this way the $N$D Euclidean system arises as a smooth  limiting flat case that can be interpreted as a contraction in terms of the curvature parameter $\k$. 

In order to prove this result we shall explicitly construct the set of  $2N-1$ functionally independent integrals of motion for the generalized KC Hamiltonian. In particular,  $2N-2$  of them will be {\em quadratic} and provided by an $\rsl(2,\RR)$ Poisson coalgebra symmetry~\cite{BH07, BEHR} (together with the Hamiltonian), while the remaining `hidden' one is {\em quartic} in the momenta and generalizes the result of~\cite{Verrier} on ${\mathbb E}^3$ to these three $N$D classical spaces of constant curvature. In this way the short list of $N$D maximally superintegrable Hamiltonians (see~\cite{annals} an references therein) is enlarged with another instance.

The paper is organized as follows. In the next section we introduce the geometric background on which the rest of the paper will be based:  the  Poincar\'e and Beltrami phase spaces arising, respectively, as the stereographic and the central projection from a linear ambient space $\RR^{N+1}$~\cite{BH07, Doub}.  The next section is devoted to recall the description of  the maximal integrability of the curved KC system in terms of these two phase spaces. In section 4  the spherical  and `hidden' nonlinear symmetries of the KC system are fully described, thus providing a detailed explanation of the techniques making possible the `transition' from the integrability properties of the curved KC system to the generalized one.  The core of the paper is contained in section 5, where we explicitly show how the spherical symmetry breaking induced by the centrifugal terms can be appropriately replaced  by  an $\rsl(2,\RR)$ Poisson coalgebra symmetry~\cite{BR, CRMAngel} that allows us to construct the `additional'   {\em quartic}  integral of motion  for the 
  $N$D  generalized  curved  KC systems. Finally, some remarks close the paper.
 

\sect{Poincar\'e and Beltrami phase spaces}

To start with we present the structure of the Poincar\'e and Beltrami phase spaces~\cite{BH07}, which will allow us to deal   with the three classical Riemannian spaces in a unified setting. Furthermore, as a new result,   we also present the explicit canonical transformation relating both of them.

Given a constant sectional curvature   $\k$,   the $N$D
spherical $\mathbb S^N$
$(\k>0)$, Euclidean $\mathbb E^N$ ($\k=0$) and hyperbolic  $\mathbb H^N$
$(\k<0)$  spaces
 can be simultaneously embedded in an ambient linear space $\mathbb R^{N+1}$ with
ambient (or Weiertrass) coordinates $(x_0,\>x)=(x_0,x_1,\dots,x_N)$ by requiring them to  fulfill  the `sphere'
constraint $\Sigma$: $x_0^2+\k\,  \>x^2=1$. Hereafter for any two $N$D vectors, say $\>a=(a_1,\dots,a_N)$ and $\>b=(b_1,\dots, b_N)$, we denote  $\>a^2=\sum_{i=1}^N a_i^2$, $|\>a|=\sqrt{\>a^2}$ and $\>a\cdot \>b =\sum_{i=1}^N a_i b_i$.
 The metric on these three Riemannian   spaces of constant curvature  is   given, in ambient coordinates, 
by~\cite{BHSS03}:
\be
\dd
s^2= {1\over\k}
\left(\dd x_0^2+\k\,  \dd \>x^2\right)\biggr|_{\Sigma}  .
\label{ba}
\ee
Now, if  we  consider the stereographic projection~\cite{Doub}  from  
$(x_0,\>x)\in
\Sigma\subset \mathbb R^{N+1}$ to the {\em Poincar\'e coordinates}
$\>\y\in \mathbb R^N$ with pole  
$(-1,\>0)\in \mathbb R^{N+1}$, that is, 
$(-1,\>0)+\la\, (1,\>\y)\in\Sigma$,    we obtain  that
\begin{equation}
\la=\frac{2}{1+  \k\>\y^2}
\qquad x_0=\la -1=\frac {1-\k\>\y^2}{1+
\k\>\y^2}\qquad 
\>x=\la\,\>\y=\frac{2\>\y}{1+
\k\>\y^2}.
\label{bb}
\end{equation}
On the other hand, if we apply the central projection   from   $ (x_0,\>x)$
to the {\em Beltrami coordinates}  $\>\ji\in \mathbb R^N$  with pole  
$(0,\>0)\in \mathbb R^{N+1}$, such that
$(0,\>0)+\m\,
(1,\>\ji)\in\Sigma$, 
 we find   
\begin{equation}
\m=\frac{1}{\sqrt{1+ \k \>\ji^2}}\qquad
 x_0=\m \qquad 
\>x=\m\, \>\ji=\frac{\>\ji}{\sqrt{1+ \k \>\ji^2}}.
\label{bc}
\end{equation}
The image of these  projections     is the subset of   $\mathbb  R^N$   determined by either $\la>0$ (and then  $  1+\k
\>q^2>0$), or by $\mu\in \RR$ (and thus $1+ \k \>\ji^2>0$). This means that for  $\mathbb S^{N}$ ($\k>0$) both projections lead to $\mathbb  R^N$ with the exception of a single point; for   $\mathbb H^N$  they give the open subset $\>q^2<1/|\k|$ or $\>\ji^2<1/|\k|$ (the Poincar\'e disc in 2D); and in both cases, if   $\k=0$, we recover $\mathbb E^N$ in     Cartesian
coordinates $\>x\in \mathbb  R^N\equiv \mathbb E^N$  since  $\>x=2\>q= \>\ji$.

Therefore, it can be shown that the metric (\ref{ba}) in both coordinate systems reads
\begin{equation}
\dd s^2=4\,\frac{\dd \>\y^2}{(1+\k\>\y^2)^2}=  
\frac{(1+\k\>\ji^2)\dd\>\ji^2-\k
(\>\ji\cdot \dd\>\ji)^2}{(1+\k\>\ji^2)^2} 
\label{bd}
\end{equation}
 so that    the corresponding geodesic flow on these spaces can be described through the free Lagrangian given (up to a positive multiplicative  constant) by:
\be
{\cal T}= 2\, \frac{\dot{\>\y}^2}{(1+\k\>\y^2)^2}  =
\frac{(1+\k\>\ji^2)\dot{\>\ji}^2-\k({\>\ji} \cdot \dot{\>\ji})^2}{2(1+\k\>\ji^2)^2} .
\label{be}
\ee
Hence  the canonical momenta $\>p$, $\>\pji$ conjugate  to $\>q$, $\>\ji$ are obtained through a Legendre
transformation yielding
\begin{equation}
\>p= 4\, \frac{\dot{\>\y}}{(1+\k\>\y^2)^2} \qquad  \>\pji=    \frac{(1+\k\>\ji^2)\dot{\>\ji}-\k (\>\ji\cdot
\dot{\>\ji})\>\ji}{(1+\k\>\ji^2)^2}   
\label{bf}
\end{equation}
and the geodesic flow kinetic energy is found to be
\be
 {\cal T}= \frac{1}{8}\left( 1+\k \>q^2\right)^2 \>p^2  =
\frac{1}{2}(1+\k \>\ji^2)\left( \>\pji^2+\k (\>\ji\cdot \>\pji)^2 \right).
\label{bg}
\ee
 Evidently, $\cal T$ can be written with a factor $1/2$ instead of $1/8$  in the Poincar\'e phase space but we have kept the latter factor in order  to make explicit the equality between Poincar\'e and Beltrami expressions.

The canonical equivalence between both phase spaces is characterized by the following statement, that can be proven through  direct computations and by taking into account the expressions (\ref{bb})--(\ref{bf}).

\medskip

\noindent
{\bf Proposition 1.} {\em Let $(\>q,\>p)$ be the Poincar\'e phase space variables such that $\{ q_i,p_j\} =\delta_{ij}$ and $(\>\ji,\>\pji)$ the Beltrami ones satisfying 
$\{\ji_i,\pji_j\}=\delta_{ij}$. Both sets of canonical variables are related through the canonical transformation given by
\be
\begin{array}{ll}
\displaystyle{   \>q=\frac{\>\ji}{ 1+\sqrt{1+\k \>\ji^2 }   } }&\quad \displaystyle{\>p=\left( 1+\sqrt{1+\k \>\ji^2 }  \right)\>\pji +\k \>\ji (\>\ji\cdot \>\pji)  }\\[10pt]
\displaystyle{ \>\ji=\frac{2 \>q}{1-\k \>q^2} }& \quad \displaystyle{ \>\pji=\frac{1-\k \>q^2}{2(1+\k \>q^2) }\left( (1+\k \>q^2)\>p- 2\k \>q (\>q\cdot\>p)  \right) } .
\end{array}
\label{bh}
\ee
}
\medskip

 Moreover, 
from (\ref{bh}) we obtain  the following useful relations to be considered below:
\bea
&&q_i p_j- q_j p_i=\ji_i \pji_j -\ji_j \pji_i\qquad \frac{q_i}{q_j}=\frac{\ji_i}{\ji_j}\qquad  
\frac{q_i}{\sqrt{\>q^2}}=\frac{\ji_i}{\sqrt{\>\ji^2}}
\nonumber\\[2pt]
&& \>\pji+\k (\>\ji\cdot \>\pji)\>\ji=\frac 12 (1-\k\bq^2) \>p+\k (\bq\cdot \bp) \>q 
\nonumber\\[2pt]
&&
\>\ji\cdot\>\pji = \frac{1-\k \>q^2}{1+\k \>q^2}\, (\bq\cdot \>p)
.
\label{bi}
\eea


\sect{The    Kepler--Coulomb  system}

In order to construct the KC Hamiltonian we recall that for the three Riemannian spaces with constant curvature, the geodesic radial distance $r$  (along the geodesic that joins the particle and the origin in the space) can be expressed, in this order,  in  ambient, Poincar\'e and Beltrami 
coordinates as~\cite{BH07}:
\begin{equation}
\frac{1}{\k}\tan^2(\sqrt{\k}\,r) =\frac{\>x^2}{x_0^2}   =\frac{4\>\y^2}{(1-\k\>\y^2)^2}=\>\ji^2.
\label{ca}
\end{equation} 
In fact, these three expressions provide the appropriate definition  for the curved (Higgs) oscillator potential~\cite{Higgs, Leemon}. 

By using these coordinates, the KC system is just the Hamiltonian~\cite{BH07}
\bea
&& {\cal H}= \frac{1}{8}\left( 1+\k \>q^2\right)^2 \>p^2  -{\kk}\, \frac{1-\k\>\y^2 }{2\sqrt{\>\y^2}}\nonumber\\
&&\quad\, 
= \frac{1}{2}(1+\k \>\ji^2)\left( \>\pji^2+\k (\>\ji\cdot \>\pji)^2 \right) -\frac{\kk}{\sqrt{\>\ji^2}}
\label{cb}
\eea
where the first term is the kinetic energy (\ref{bg})  and the second one is the curved KC potential, which is obtained as the square root of the inverse of (\ref{ca}). We stress that under this framekork we are able to cover simultaneously  the three cases (\ref{ab}) for the particular cases of the sectional curvature  $\k\in \{+1, 0, -1\}$. Moreover, in this language the limit $\k\to 0$ corresponds to  the (flat) contraction  $\mathbb S^N\rightarrow   \mathbb E^N \leftarrow\mathbb H^N$. 

The maximal superintegrability of  the curved KC Hamiltonian (\ref{cb})  is characterized by

\medskip

\noindent
{\bf Proposition 2.}~\cite{BH07} {\em Let $\cal H$ be the KC Hamiltonian (\ref{cb}) and let us consider the quadratic  functions in the momenta given by
\bea
&& C^{(m)}= \sum_{1\leq i<j}^m  ({q_i}{p_j} -
{q_j}{p_i})^2  \qquad\qquad 
  C_{(m)}= \sum_{N-m+1\leq i<j}^N   ({q_i}{p_j} -
{q_j}{p_i})^2  
\label{cc}\\
&&\displaystyle{ {\cal L}_i =\sum_{l=1 }^N \left(
\frac 12 (1-\k
\>q^2) p_l + \k (\>q\cdot
\>p) q_l 
\right) (q_l p_i-q_i p_l) 
+\kk\, \frac{ q_i}{\sqrt{\>q^2}}   }\nonumber\\
&&  \quad\, =\sum_{l=1 }^N  \left(
\pji_l+\k (\>\ji\cdot \>\pji) \ji_l \right)   (\ji_l \pji_i-\ji_i \pji_l)  
 + \kk\, \frac{ \ji_i}{\sqrt{\>\ji^2}}  .
\label{cd}
\eea
where  $m=2,\dots, N$,  $C^{(N)}=C_{(N)}$ and $i=1,\dots,N$. 
Then:\\
(i)  The $2N-3$ functions (\ref{cc}) and  the $N$ functions (\ref{cd}) Poisson-commute with $\cal H$.\\
(ii)  Each set 
$\{{\cal H},C^{(m)}\}$ and $\{{\cal H},C_{(m)}\}$ $(m=2,\dots, N)$ provides $N$ functionally independent functions   in
involution. \\
(iii) For a fixed $i$, the $2N-1$ functions $\{ {\cal H}, C^{(m)}, C_{(m)} ,  {\cal L}_i \}$ with $m=2,\dots, N$   are functionally independent.
 }
\medskip

In order to illustrate this result, we point out that that the   Euclidean KC system is obtained from proposition 2 by setting $\k=0$. In this case, the set of integrals $C^{(m)}$ and $C_{(m)}$ do not change, as they reflect the `abstract' spherical symmetry of the system. On the contrary, the ${\cal L}_i$ integral reads
\be
 {\cal L}_i =\sum_{l=1 }^N 
\frac 12  p_l  (q_l p_i-q_i p_l) 
+\kk\, \frac{ q_i}{\sqrt{\>q^2}}  =
\sum_{l=1 }^N  
\pji_l  (\ji_l \pji_i-\ji_i \pji_l)  
 + \kk\, \frac{ \ji_i}{\sqrt{\>\ji^2}}  .
\label{fff}
\ee
Of course, this is indeed the well known expression for the $i$-th component of the Laplace--Runge--Lenz vector on $ \mathbb E^N$; note that the only difference between both expressions is the $1/2$ factor for the Poincar\'e variables   since   $\>q=\frac 12 \>\ji$ and $\>p=2\>\pji$ when $\k=0$  as it  follows from  (\ref{bh}).


 \sect{Symmetries of the Kepler--Coulomb  system}

At this point, a detailed analysis of the the symmetry properties of the integrals of the motion for the KC Hamiltonian is worth to be performed, since this background will provide the appropriate framework to extend such integrability properties to the generalized KC system.

 \subsect{Spherical symmetry}

Firstly, we stress that the constants of the motion
 (\ref{cc})  keep the same form in both Poincar\'e and Beltrami phase spaces (see (\ref{bi})). They reflect the well known spherical symmetry of the KC system (and of any central potential as well).  In particular, 
the functions $J_{ij}={q_i}{p_j} - {q_j}{p_i}$ with $i<j$ and $i,j=1,\dots,N$ span an $\frak
{so}(N)$ Lie--Poisson algebra  
\begin{equation}
\{ J_{ij},J_{ik} \}= J_{jk} \qquad  \{ J_{ij},J_{jk} \}= -J_{ik} \qquad 
\{ J_{ik},J_{jk} \}= J_{ij} \qquad i<j<k .
\label{ce}
\end{equation}
Thus   the constants of the motion (\ref{cc})  are the Casimirs of certain rotation subalgebras
$\frak {so}(m)\subset \frak {so}(N)$ written through the sums of the square of  {\em some} angular momentum components $J_{ij}$. The square of the total angular momentum $\bjj^2$ is then given by the Casimir of 
$ \frak {so}(N)$:
\be
\bjj^2=C^{(N)}=C_{(N)}=  \sum_{1\leq i<j}^N  J_{ij}^2 .
\label{cf}
\ee

This    $\frak {so}(N)$-symmetry can be further enlarged by defining the following $N$   functions $P_i$ $(i=1,\dots,N)$
\be
P_i= 
\frac 12 (1-\k
\>q^2) p_i + \k (\>q\cdot
\>p) q_i
= \pji_i+\k (\>\ji\cdot \>\pji) \ji_i  ,
\label{ccff}
\ee
which come from the first factors of the integrals ${\cal L}_i$  (\ref{cd}) (so these are specifically characterized by the proper KC system). Their Lie--Poisson brackets  are given by
\be
\{ J_{ij},  P_k \}=\delta_{ik}  P_j - \delta_{jk}  P_i
\qquad \{ P_i,P_j\}=\k J_{ij} .
\label{ccgg}
\ee
Therefore, the  $N(N+1)/2$ functions $\langle J_{ij}, P_i \rangle$ $(i<j;\ i,j=1,\dots,N)$ span  a Lie--Poisson algebra   $\frak
{so}_\k(N+1)$~\cite{vulpiPAN, BHSS03},  with commutation relations given by (\ref{ce})  and (\ref{ccgg}), in which $\k$ can be interpreted as a  contraction parameter. These Poisson brakets define
the spherical $\frak
{so}(N+1)$ for $\k>0$,    hyperbolic $\frak
{so}(N,1)$ for $\k<0$, and the  Euclidean Lie--Poisson algebra $\frak
{iso}(N)$ for $\k=0$.   

Moreover, the three  Riemannian spaces of  {constant sectional curvature }$\k$, described in section 2, can be constructed as homogeneous spaces through the quotient $\langle J_{ij}, P_i \rangle/ \langle J_{ij}  \rangle=\langle P_i  \rangle$:
\be
\begin{array}{ll}
 \mathbb S^{N} = \frak
{so}(N+1)/   \frak
{so}(N)    &\ {\mbox { for}}\ \k>0 ; \\[2pt]
\mathbb E^{N} = \frak
{iso}(N)/   \frak
{so}(N)   &\ {\mbox { for}}\ \k=0  ;\\[2pt]
\mathbb H^{N} = \frak
{so}(N,1)/  \frak
{so}(N)  &\ {\mbox { for}}\ \k<0  ;
\end{array}
\label{cj}
\ee
with the   $P_i $'s playing the role of  (curved) translations on such spaces, and the contraction $\k=0$ gives rise to the Euclidean (commutative) translations $P_i=\frac 12 p_i=\pji_i$.
The Poisson brackets between the Hamiltonian $\cal H$ (\ref{cb}) and the $\frak
{so}_\k(N+1)$ generators read
\be
\{ {\cal H}, J_{ij}\}=0\qquad \{ {\cal H}, P_i\}=\kk\, \frac{q_i(1+\k \>q^2)^2}{4|\>q|^3}  = \kk\, \frac{\ji_i(1+\k \>\ji^2)}{|\>\ji|^3}\equiv \kk\, \frac{x_i}{\>x^3}.
\label{cdj}
\ee

We stress that from the viewpoint of such $\frak
{so}_\k(N+1)$-symmetry, the constants of the motion (\ref{cd})  can be expressed in a very natural and simple form, as~\cite{vulpiPAN}  
 \be{ {\cal L}_i =\sum_{l=1 }^N  P_l  J_{li}
+\kk\, \frac{ q_i}{\sqrt{\>q^2}}    } 
\label{ccjj}
 \ee
 such that $J_{ii}\equiv 0$ and $J_{li}=-J_{il}$ if $l>i$. This, in turn, directly  shows the functionally independence of a given $ {\cal L}_i$ with respect to the integrals (\ref{cc}), as stated in proposition 2,  since the latter  are only formed by rotation generatos $J_{ij}$ of $\frak {so}(N)$.  
 

 \subsect{Nonlinear  angular momentum symmetry}

By taking into account (\ref{ccjj}),   it is a matter of straightforward computations to show that  the $N$ constants of the motion (\ref{cd}) are transformed as an $N$-vector under the $\frak
{so}(N)$ generators (\ref{ce}),
\be
\{ J_{ij},  {\cal L}_k \}=\delta_{ik}  {\cal L}_j - \delta_{jk}  {\cal L}_i .
\label{cg}
\ee
The $N$ functions ${\cal L}_i$  are, in fact,  the components of  the {\em Laplace--Runge--Lenz  $N$-vector} on  $\mathbb S^{N}$,  $\mathbb H^{N}$ and  $\mathbb E^{N}$, which correspond to the `hidden' symmetries of the KC Hamiltonian.  Moreover,  the Lie--Poisson brackets involving the  ${\cal L}_i$ components  read
\be
\{ {\cal L}_i, {\cal L}_j\} = 2 \left(\k \bjj^2 - {\cal H} \right) J_{ij} \equiv \Lambda J_{ij}  .
\label{ch}
\ee
This expression is worth to be compared with (\ref{ccgg}). 
Since   $ {\cal H}$ Poisson-commutes with all the functions $J_{ij}$ and    $ {\cal L}_i$ (in algebraic terms we would say that  $ {\cal H}$ behaves as   a central extension),   we find that the set of $N(N+1)/2$ functions $\langle J_{ij}, {\cal L}_i \rangle$ $(i<j;\ i,j=1,\dots,N)$ span  a {\em nonlinear} (cubic) Poisson algebra that we sall denote as $\frak
{so}_\k^{(3)}(N+1)$.  Only in $\mathbb E^{N}$ ($\k=0$) this nonlinear symmetry algebra reduces to    Lie--Poisson algebras with $\Lambda=-2\cal H$ being a constant, and we get  $\frak{so}(N+1)$ for ${\cal H}<0$ or $\frak{so}(N,1)$  for ${\cal H}>0$.

Therefore, from a geometrical viewpoint we can say that the role of the translations $P_i$ on the homogeneous spaces (\ref{cj}), with Lie--Poisson symmetry $\frak
{so}_\k (N+1)$, is replaced by the Laplace--Runge--Lenz components $ {\cal L}_i$, with nonlinear symmetry $\frak
{so}_\k^{(3)}(N+1)$, whereas the role of the former constant curvature $\k$ is now played by the  {\em quadratic function} $\Lambda$ (\ref{ch}). In this respect, notice that although $\bjj^2$ and ${\cal H}$ are both constants of the motion for the KC system, $\bjj^2$ does not Poisson-commute with $ {\cal L}_i$. As a consequence,  $\Lambda$ is not a central function within  $\frak{so}_\k^{(3)}(N+1)$.

However, in the case $N=2$ we have that  $\>J^2\equiv J_{12}^2$,  so that $\frak
{so}_\k^{(3)}(3)=\langle J_{12},{\cal L}_1,{\cal L}_2\rangle$ gives rise to the
 {\em cubic}     Poisson algebra 
\be 
\{ J_{12}, {\cal L}_1\}= {\cal L}_2\qquad \{ J_{12}, {\cal L}_2 \}=-{\cal L}_1
\qquad \{ {\cal L}_1, {\cal L}_2 \}=2\k J_{12}^3 -2 {\cal H} J_{12} .
\label{ck}
\ee 
 Next we can define   `number' ${\cal L}$ and `ladder' operators ${\cal L}_\pm$ as
$$
{\cal L}={\rm i} J_{12}\qquad {\cal L}_\pm = {\cal L}_1\pm {\rm i} {\cal L}_2
 \label{cm}
$$
   fulfilling the cubic  commutation relations 
\be
\{ {\cal L},{\cal L}_\pm\}=\pm {\cal L}_\pm\qquad \{ {\cal L}_+ , {\cal L}_-\}= 4 \k {\cal L}^3 +4 {\cal H} {\cal L}
\ee
which reproduce the  Poisson algebra analogue of the  Higgs
 $\frak
{sl}^{(3)}(2,\mathbb R)$    algebra~\cite{Higgs} whenever $\k\ne 0$. In this case the Poisson brackets are associated to $ \mathbb S^{2}$ and $ \mathbb H^{2}$, whereas  the contraction $\k=0$ gives  $\frak
{sl}(2,\mathbb R)$  (or  $\frak
{gl}(2)$ if $ {\cal H}$ is considered as an actual generator) for $ \mathbb E^{2}$. Note that the Higgs algebra is endowed with a {\em quartic} Casimir function given by
\be
{\cal C}_{\frak
{sl}^{(3)}(2,\mathbb R)}={\cal L}_+ {\cal L}_- +\k {\cal L}^4 + 2  {\cal H}   {\cal L}^2 . 
\ee
Moreover, if we realize this Casimir  in terms of Poincar\'e or Beltrami coordinates, we obtain that ${\cal C}_{\frak
{sl}^{(3)}(2,\mathbb R)}=\kk^2$, which is just the square of the coupling constant of the KC potential.  It is worth to stress that the Higgs algebra  has been deeply studied and applied to different  quantum physical  models (beyond integrable systems) with an underlying nonlinear angular momentum symmetry (see~\cite{r1, ref6, ref6b,   r3, ref6c, ref7, osvaldo} and references therein).


\sect{The generalized Kepler--Coulomb  system}

The generalized KC  Hamiltonian ${\cal H}^\g$ is obtained by adding  $N$ centrifugal terms (with non-vanishing parameters $b_i\in \RR$) to the KC system ${\cal H}$ (\ref{cb}). In the curved cases, the way to define appropriately such centrifugal terms was presented and fully explained in~\cite{BH07}. Explicitly,  in Poincar\'e coordinates the generalized KC system is given by
\bea
&&{\cal H}^\g={\cal H}+\frac 18 \left(1+\k \>q^2\right)^2\sum_{i=1}^N\frac{b_i}{q_i^2}\nonumber\\
&&\quad \ \,=
\frac{1}{8}\left( 1+\k \>q^2\right)^2 \>p^2  -{\kk}\, \frac{1-\k\>\y^2 }{2\sqrt{\>\y^2}}
+\frac 18 \left(1+\k \>q^2\right)^2\sum_{i=1}^N\frac{b_i}{q_i^2}
\label{dcy}
\eea
while in terms of Beltrami variables this reads
\bea
&&{\cal H}^\g={\cal H}+\frac 12 (1+\k \>\ji^2)\sum_{i=1}^N\frac{b_i}{\ji_i^2}\nonumber\\
&&\quad \ \,= \frac{1}{2}(1+\k \>\ji^2)\left( \>\pji^2+\k (\>\ji\cdot \>\pji)^2 \right) -\frac{\kk}{\sqrt{\>\ji^2}}+\frac 12 (1+\k \>\ji^2)\sum_{i=1}^N\frac{b_i}{\ji_i^2}.
\label{dcz}
\eea
This is the curved $\k$-analogue of the generalized KC system on  ${\mathbb E}^N$  (\ref{aa}). We recall that  the centrifugal terms  are proper centrifugal barriers on both ${\mathbb E}^N$ ($\k=0$) and ${\mathbb H}^N$ ($\k<0$). Moreover, only on ${\mathbb S}^N$ ($\k>0$) all these terms can be alternatively interpreted as non-central harmonic oscillators   (see~\cite{kiev,BHSS03,ran1, ran2} for a full discussion on the subject).

The maximal superintegrability of this Hamiltonian, which constitute the main result of this paper, can now  be stated and proven as follows.

\medskip

\noindent
{\bf Theorem.} {\em Let ${\cal H}^\g$ be the  generalized KC Hamiltonian (\ref{dcy})--(\ref{dcz}) with all  $b_i\ne 0$.  Let us consider the quadratic and quartic functions in the momenta given by
\bea
&& C^{(m)}_\g= \sum_{1\leq i<j}^m \left\{ ({q_i}{p_j} -
{q_j}{p_i})^2 + \left(
b_i\frac{q_j^2}{q_i^2}+b_j\frac{q_i^2}{q_j^2}\right)\right\}
+\sum_{i=1}^m b_i \nonumber\\
&& C_{(m)}^\g= \sum_{N-m+1\leq i<j}^N \left\{ ({q_i}{p_j} -
{q_j}{p_i})^2 + \left(
b_i\frac{q_j^2}{q_i^2}+b_j\frac{q_i^2}{q_j^2}\right)\right\}
+\sum_{i=N-m+1}^N b_i  \label{db}  
\eea
\bea
&&\!\!\!\!\!  \llkc_i =\left(    \sum_{l=1 }^N \left(
\frac 12 (1-\k
\>q^2) p_l + \k (\>q\cdot
\>p) q_l 
\right) (q_l p_i-q_i p_l) 
+ \frac{\kk q_i}{\sqrt{\>q^2}} - (1-\k \>q^2)\sum_{l=1}^N \frac{b_l  q_i }{2 q_l^2}\right)^2 \nonumber\\
&&\qquad\quad+ \frac{b_i}{q_i^2} \left(  \sum_{l=1}^N \left( \frac 12 (1-\k
\>q^2) p_l + \k (\>q\cdot
\>p) q_l  \right) q_l  \right)^2 \nonumber\\
&&    \  \  =   \left( \sum_{l=1 }^N  \left(
\pji_l+\k (\>\ji\cdot \>\pji) \ji_l \right)   (\ji_l \pji_i-\ji_i \pji_l)  
 + \frac{\kk \ji_i}{\sqrt{\>\ji^2}}  -\sum_{l=1}^N\frac{b_l}{\ji_l^2}\,\ji_i \right)^2 \nonumber\\
 &&\qquad\quad +  \frac{b_i}{\ji_i^2} \left(  \sum_{l=1}^N \left(  \pji_l+\k (\>\ji\cdot \>\pji) \ji_l \right) \ji_l  \right)^2
\label{dc}
\eea
where $m=2,\dots, N$,   $C^{(N)}_\g=C_{(N)}^\g$  and  $i=1,\dots,N$.   Then:\\
(i)  The $2N-3$ functions (\ref{db}) and   the $N$ functions (\ref{dc}) Poisson-commute with ${\cal H}^\g$.\\
(ii)  Each set 
$\{{\cal H}^\g,C^{(m)}_\g\}$ and $\{{\cal H}^\g,C_{(m)}^\g\}$ $(m=2,\dots, N)$ provides $N$ functionally independent functions   in
involution. \\
(iii) For a fixed $i$, the $2N-1$ functions $\{ {\cal H}^\g, C^{(m)}_\g, C_{(m)}^\g,  \llkc_i\}$  with $m=2,\dots, N$ are functionally independent.
 }
\medskip

\noindent{\bf Proof}. We shall proceed in two steps. Firstly, we shall prove that: (1) ${\cal H}^\g$     Poisson-commutes with all the integrals $C^{(m)}_\g$ and $C_{(m)}^\g$ (\ref{db}) and (2) that  the sets $\{{\cal H}^\g,C_\g^{(m)}\}$ and 
$\{{\cal H}^\g,C^\g_{(m)}\}$ ($m=2,\dots,N$) are formed by $N$ functionally independent functions in involution.

These two statements can be immediately proven by making use of the $\mathfrak{sl}(2,\RR)$-coalgebra symmetry~\cite{BH07,BEHR,BR, CRMAngel} of the generalized KC Hamiltonian. Indeed, this coalgebra symmetry is just the appropriate generalization of the spherical symmetry for this problem. In particular, let us define the functions
\be
J_-=\bq^2 \qquad J_3=\bq\cdot\bp \qquad J_+=\bp^2+\sum_{i =1}^N \frac{b_i }
{q_i^{2}} \label{Jep}
\ee
where $\bq$ and $\bp$ are,  in principle,  `abstract' canonical variables (in this paper they can be directly identified with  either Poincar\'e or Beltrami ones). These three functions span the $\rsl(2,\mathbb R)$ Poisson  coalgebra with Poisson brackets, coproduct and Casimir function given by
\bea
&&\{J_3,J_+\}=2J_+ \qquad \{J_3,J_-\}=-2J_- \qquad
\{J_-,J_+\}=4J_3 \nonumber \\
&&\Delta(J_l)=J_l\otimes 1 + 1 \otimes J_l\qquad l=+,-,3\nonumber\\
&& {\cal C}= J_- J_+ - J_3^2.\label{jpc}
\eea
We say that the generalized KC Hamiltonian is endowed with this symmetry because it can be written
  in terms of the $\mathfrak{sl}(2,\RR)$-coalgebra generators, either   in Poincar\'e variables  (\ref{dcy}), as
\be
{\cal H}^\g=   \frac{1}{8}\left( 1+\k J_-\right)^2 J_+  -{\kk}\, \frac{1-\k J_- }{2\sqrt{J_-}} 
\label{ma}
\ee
or in Beltrami ones (\ref{dcz}) as
\be
{\cal H}^\g= 
\frac{1}{2}(1+\k J_-)\left( J_++\k  J_3^2 \right) -\frac{\kk}{\sqrt{J_-}} .
\label{mbb}
\ee
This, in turn, shows that the appearance of centrifugal terms is deeply related with the $\mathfrak{sl}(2,\RR)$ symplectic realization   (\ref{Jep}) through the generator $J_+$. Once this symmetry for ${\cal H}^\g$ has been proven, the statements (1) and (2) are a direct consequence of such coalgebra invariance (see~\cite{BH07,BEHR, CRMAngel} for a detailed explanation). In particular, ${\cal H}^\g$  (\ref{ma})   Poisson-commutes with all the integrals $C^{(m)}_\g$ and $C_{(m)}^\g$ (\ref{db}) since the latter are just the left
and right $m$th-coproducts of the Casimir   (\ref{jpc}), respectively, under the symplectic realization (\ref{Jep}). On the other hand,  the sets $\{{\cal H}^\g,C_\g^{(m)}\}$ and 
$\{{\cal H}^\g,C^\g_{(m)}\}$ ($m=2,\dots,N$) are, by construction, given  by $N$ functionally independent integrals of the motion in involution. 

Now the second task to complete the proof concerns the `additional' quartic integral $\llkc_i$ (\ref{dc}). The proof that this  Poisson-commutes with  ${\cal H}^\g$ can be achieved through    direct but cumbersome computations, that can be carried out  by starting with the expression (\ref{dcy}) (or (\ref{dcz})) and by using   the results  given in the   previous section. Finally, the functional independence of $\llkc_i$ with respect to the set  $\{ {\cal H}^\g, C^{(m)}_\g, C_{(m)}^\g \}$ is automatically fulfilled by considering all these quantities as $N$-parametric smooth deformations  in   $\>b=(b_1,\dots,b_N)$ of the corresponding non-generalized objects, {\em i.e.}:
\be
{\cal H^\g}= {\cal H}+o({\bf b})\quad
C^{(m)}_\g = C^{(m)} +o({\bf b})\quad
C_{(m)}^\g = C_{(m)} +o({\bf b})\quad
\llkc_i={\cal L}_i + o({\bf b}) .
\ee
Hence the proof of the theorem follows by taking into account proposition 2, since we know that ${\cal L}_i $ is  functionally independent with respect to the set $\{ {\cal H}, C^{(m)}, C_{(m)} \}$.
\hfill $\square$

\bigskip

Some remarks are in order:

\begin{itemize}

\item The constants of the motion (\ref{db}) again keep the same form both in Poincar\'e and Beltrami variables due to the relations (\ref{bi}). 

\item The `hidden' constants of the motion $ \llkc_i$ (which are quartic functions in the momenta) can be written in terms of the rotation generators $J_{ij}$ (\ref{ce}), the translation ones $P_i$ (\ref{ccff}) and  the components of the Laplace--Runge--Lenz $N$-vector ${\cal L}_i$ (\ref{cd}) associated to the KC system. Namely,
\bea
&&\!\!\!\!\!  \llkc_i =\left(    \sum_{l=1 }^N P_l J_{li}
+ \frac{\kk q_i}{\sqrt{\>q^2}} - (1-\k \>q^2)\sum_{l=1}^N \frac{b_l  q_i }{2 q_l^2}\right)^2 + \frac{b_i}{q_i^2} \left(  \sum_{l=1}^N  P_l q_l  \right)^2 \nonumber\\
&&    \   =\left(     {\cal L}_i - (1-\k \>q^2)\sum_{l=1}^N \frac{b_l  q_i }{2 q_l^2}\right)^2 + \frac{b_i}{q_i^2} \left(  \sum_{l=1}^N  P_l q_l  \right)^2 .
\label{dcx}
\eea

\item The spherical symmetry breaking due to centrifugal terms implies that $\{ {\cal H}^\g,J_{ij}\}\ne 0$ and that $\>J^2$ (\ref{cf}) is no longer a constant of motion (its role is now played by $C^{(N)}_\g=C_{(N)}^\g$). In fact, the Poisson brackets (\ref{cdj}) are now generalized as
\bea
&& \{ {\cal H}^\g, J_{ij}\}=\frac{b_i q_j^4 - b_j q_i^4}{4q_i^3 q_j^3}(1+\k \>q^2)^2=\frac{b_i \ji_j^4 - b_j \ji_i^4}{\ji_i^3 \ji_j^3}(1+\k \>\ji^2)\label{cdjj} \\
&& \{ {\cal H}^\g, P_i\} = \frac{ 2  \kk q_i^4 - b_i |\>q|^3(1-\k \>q^2)}{8 q_i^3|\>q|^3}    (1+\k \>q^2)^2  = \frac{   \kk \ji_i^4 - b_i |\>\ji|^3}{\ji_i^3|\>\ji|^3}    (1+\k \>\ji^2). 
\nonumber
\eea

\end{itemize}

\bigskip

As a byproduct of the above results we straightforwardly recover  the quasi-generalized KC  systems~\cite{BH07}:
\medskip

\noindent
{\bf Corollary.} {\em   Let ${\cal H}^\g_i$ be the quasi-generalized KC Hamiltonian (\ref{dcy})--(\ref{dcz}) with a single  $b_i= 0$   (so the index $i$ is fixed) and let us consider the   quadratic  functions in the momenta (\ref{db}) and\bea
&&\displaystyle{ \lkc_i =\sum_{l=1 }^N \left(
\frac 12 (1-\k
\>q^2) p_l + \k (\>q\cdot
\>p) q_l 
\right) (q_l p_i-q_i p_l) }\nonumber\\
&&\qquad\qquad 
+\kk\, \frac{ q_i}{\sqrt{\>q^2}} - (1-\k \>q^2)\sum_{l=1;l\ne i}^N \frac{b_l}{2 q_l^2}\, q_i  \nonumber\\
&&   \qquad\! =\sum_{l=1 }^N  \left(
\pji_l+\k (\>\ji\cdot \>\pji) \ji_l \right)   (\ji_l \pji_i-\ji_i \pji_l)  
 + \kk\, \frac{ \ji_i}{\sqrt{\>\ji^2}}  -\sum_{l=1;l\ne i}^N\frac{b_l}{\ji_l^2}\,\ji_i .
\label{dccc}
\eea
Then:\\
(i)  ${\cal H}^\g_i$  Poisson-commutes with the $2N-3$ functions (\ref{db}) and    the $N$  functions (\ref{dccc}).\\
(ii)  Each set 
$\{{\cal H}^\g_i,C^{(m)}_\g\}$ and  $\{ {\cal H}^\g_i,C_{(m)}^\g\}$ $(m=2,\dots, N)$ is formed by $N$ functionally independent functions   in
involution. \\
(iii) The $2N-1$ functions $\{{\cal H}^\g_i, C^{(m)}_\g, C_{(m)}^\g ,  \lkc_i \}$ with $m=2,\dots, N$    are functionally independent.
 }
\medskip

  Obviously, if a second parameter  $b_j$ also vanishes we obtain an additional integral of motion   $ \lkc_{j}$  for the Hamiltonian (that we can now call ${\cal H}_{ij}^\g$).   In that case we have
  \be
 \{ {\cal H}_{ij}^\g, J_{ij} \}= \{ {\cal H}_{ij}^\g,  \lkc_{i} \}=\{ {\cal H}_{ij}^\g,  \lkc_{j} \}=0 ,
\ee
and the rotation generator $J_{ij}$ becomes a constant of the motion (see (\ref{cdjj})).
Moreover, the three functions $\langle J_{ij}, \lkc_{i},\lkc_{j} \rangle$ fulfill the Poisson brackets
   \be
\{ J_{ij},  \lkc_{i} \}=\lkc_j\qquad  \{ J_{ij},  \lkc_{i} \}=-\lkc_j\qquad 
\{ \lkc_{i}, \lkc_{j} \} = 2 \left(\k C^{(N)}_\g - {\cal H}_{ij}^\g \right) J_{ij}
\label{xa}
\ee
  to be compared with (\ref{cg}) and (\ref{ch}). Note that these brackets   do not close a nonlinear Poisson algebra due to $C^{(N)}_\g$ (\ref{db}). The very same process follows when yet one more centrifugal parameter vanishes, and so on. From this viewpoint the KC system (\ref{cb}) arises as the  `degenerate' case   with all the $b_i=0$  of  the  generalized one (\ref{dcy})--(\ref{dcz}) in the sense that ${\cal H}^\g\to {\cal H}^\g_{12\dots N}\equiv {\cal H}$ and $ \llkc_i\to \lkc_i\to {\cal L}_i$ $(i=1,\dots,N)$.
 
Finally we would like to point out that the results here presented on the generalized KC system on curved spaces make this system `closer' to the Smorodinsky--Winternitz    system (i.e., the superposition of the curved harmonic oscillator potential (\ref{ca})  with $N$ centrifugal terms) on such spaces~\cite{kiev, Ev90, Miguel, RS, PogosClass1, PogosClass2, vulpiPAN, BHSS03, FMSUW65,Ev90b, 8,10, 11}. Both of them are maximally superintegrable, and the only structural difference between them is the fact that all the integrals of the Smorodinsky--Winternitz system are quadratic in the momenta.

\section*{Acknowledgments}

This work was partially supported by the Spanish  Ministerio de   Ciencia e Innovaci\'on  under grant  MTM2007-67389  (with EU-FEDER support) and by Junta de Castilla y
Le\'on  under project GR224. FJH is very grateful to P. Winternitz  for pointing out this problem.

\end{document}